\documentclass[conference]{IEEEtran}
\IEEEoverridecommandlockouts
\usepackage{cite}
\usepackage{amsmath,amssymb,amsfonts}
\usepackage{graphicx}
\usepackage{textcomp}
\usepackage{xcolor}
\usepackage{comment}
\usepackage{booktabs}
\usepackage{subfig}
\usepackage{amsmath}
\usepackage{array} 
\usepackage{wrapfig}
\usepackage{algpseudocode}
\algrenewcommand{\algorithmiccomment}[1]{\hfill\textbf{//}\,#1}
\usepackage{algorithm}
\algrenewcommand\alglinenumber[1]{\small #1:}
\usepackage{circledsteps}
\pgfkeys{/csteps/inner color=white}
\pgfkeys{/csteps/fill color=black}
\usepackage{pifont}
\usepackage{multirow}

\edef\oldtt{\ttdefault}
\usepackage[scaled]{beramono}
\usepackage[T1]{fontenc}
\renewcommand{\ttdefault}{\oldtt}
\newcommand{\bera}[1]{\text{\fontfamily{fvm}\selectfont #1}}
\usepackage{mathtools}

\def\BibTeX{{\rm B\kern-.05em{\sc i\kern-.025em b}\kern-.08em
    T\kern-.1667em\lower.7ex\hbox{E}\kern-.125emX}}

\usepackage{setspace}
\usepackage{eso-pic}

\begin{document}
\bstctlcite{IEEEexample:BSTcontrol}

\AddToShipoutPictureFG*{%
  \AtPageUpperLeft{%
    \makebox[\pdfpagewidth]{\raisebox{\dimexpr-\height-20pt}{%
      \large This work is accepted to the 60th Design Automation Conference (DAC), 2023
    }}%
  }%
}

\title{\huge BP-NTT: Fast and Compact in-SRAM Number Theoretic Transform with Bit-Parallel Modular Multiplication\\
}

\author{\IEEEauthorblockN{Jingyao Zhang$^*$, Mohsen Imani$^\dagger$, Elaheh Sadredini$^*$}
\IEEEauthorblockA{
\textit{$^*$University of California, Riverside $^\dagger$University of California, Irvine}\\
\{jzhan502, elahehs\}@ucr.edu, m.imani@uci.edu}
}

\maketitle
\begin{abstract}
Number Theoretic Transform (NTT) is an essential mathematical tool for computing polynomial multiplication in promising lattice-based cryptography. However, costly division operations and complex data dependencies make efficient and flexible hardware design to be challenging, especially on resource-constrained edge devices. Existing approaches either focus on only limited parameter settings or impose substantial hardware overhead. In this paper, we introduce a hardware-algorithm methodology to efficiently accelerate NTT in various settings using in-cache computing. By leveraging an optimized bit-parallel modular multiplication and introducing costless shift operations, our proposed solution provides up to 29$\times$ higher throughput-per-area and 10-138$\times$ better throughput-per-power compared to the state-of-the-art. 


\end{abstract}

\section{Introduction}\label{intro}

With the rise of cloud computing and the Internet of Things (IoT), concerns about data privacy and security are escalating, especially for vulnerable edge devices.
Lattice-based cryptography is the most promising candidate to serve as the foundation of future information security due to its superior balance of security and operational speed. Currently, three of four NIST-standardized post-quantum cryptography algorithms (PQCs) \cite{computer_security_division_selected_2017} and almost all homomorphic encryption (HE) schemes are based on lattice-based cryptography.
Typically, lattice-based cryptography is primarily based on the hardness of two problems: module learning with error (for PQCs) and ring learning with error (for HEs). The algorithms based on these two problems involve polynomial operations, such as modular addition and modular multiplication. With a complexity of $O(N^2)$, the modular multiplication of polynomials is the most time-consuming operation.

To mitigate the computing bottleneck, number-theoretic transform (NTT) is commonly employed to accelerate polynomial modular multiplication with the principles of the Fast Fourier Transform (FFT), which lowers the computing complexity of polynomial multiplication to $O(NlogN)$. 
However, the complicated data dependencies among different NTT stages and the required division operations make efficient hardware-based acceleration challenging.

To accelerate NTT effectively, ASIC/FPGA-based hardware acceleration designs are proposed \cite{ni_high-performance_2021,banerjee_sapphire_2019,xing_compact_2021}. Although performance is enhanced, they still suffer from frequent data movement between processing components and memory, which inhibits further performance growth. 
To eliminate the data movement bottleneck, in-memory computing techniques for cryptography algorithms are proposed \cite{zhang2022sealer,zhang2022inhale,wang2023infinity,li_mentt_2022,park_rm-ntt_2022,nejatollahi_cryptopim_2020}.
The challenges with existing solutions are that they (1) expand the trusted computing base to off-chip memories \cite{li_mentt_2022}, thus, introducing security vulnerabilities; (2) introduce complex peripheral circuits \cite{li_mentt_2022,park_rm-ntt_2022}, thus, incurring high area overhead;
or (3) are specialized only for NTT processing\cite{nejatollahi_cryptopim_2020}, thus, sacrificing generality and flexibility. 
These restrictions make it even more challenging to enable secure computing on vulnerable and resource-constrained edge devices.

\begin{figure}[tp]
\centerline{\includegraphics[width=3.4in]{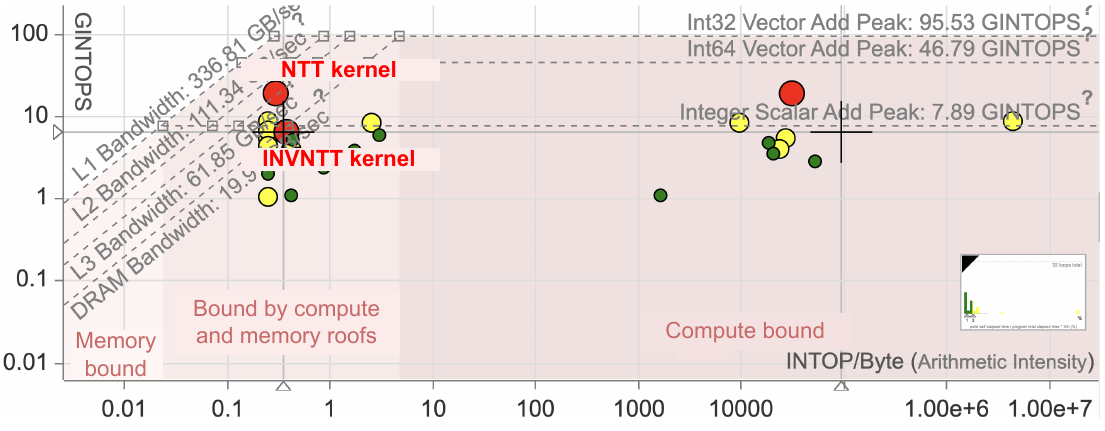}}
\caption{Roofline model for lattice-based cryptography.}
\vspace{-6mm}

\label{advisor}
\end{figure}

To analyze the computational bottleneck of the NTT, modular multiplication, and reduction (these kernels count for more the 50\% of the computation in PQC algorithms based on our profiling on CPU) and to answer the question of \textit{"where is the best place to compute in the memory hierarchy?"}, we first generate the roofline model of the lattice-based cryptography algorithms, such as CRYSTAL-Dilithium\cite{ducas_crystals-dilithium_2018} and CRYSTAL-Cyber\cite{bos_crystals-kyber_2018} using Intel Advisor \cite{noauthor_design_nodate} (Fig. \ref{advisor}). Our observation is that the main performance bottleneck for these kernels are the L1 and L2 bandwidth, and they are not bounded by the memory bandwidth bottleneck. 

Based on these insights, we re-purpose existing on-chip 6T SRAM arrays into large vector computation units and co-design them with a novel bit-parallel modular multiplication algorithm and our proposed implicit shift operations to enable energy-efficient, fast, and low-overhead NTT acceleration, especially for the IoT devices.
Our proposed solution, BP-NTT (Bit-Parallel NTT), addresses the inefficiency of existing schemes while preserving safety and flexibility. 
Since only the chip itself can be considered a trusted computing base with any off-chip data requiring encryption \cite{costan_intel_2016}, our proposed solution provides data confidentiality by not offloading the plaintext to off-chip memories.
Enabled by our proposed data organization and modular multiplication, a single 256$\times$256 SRAM subarray in BP-NTT design can support up to a 250-point polynomial with 256-bit coefficients or a 4500-point polynomial with 14-bit coefficients, which covers requirements of the lattice-based PQC algorithms (256/1024-point polynomial with 14/16/32-bit coefficients) \cite{bos_crystals-kyber_2018,ducas_crystals-dilithium_2018,fouque_falcon_nodate} and three security levels of HE under the \texttt{BKZ.qsieve} model (1024-point polynomial with 16/21/29-bit coefficients) \cite{albrecht_homomorphic_2019}.

In summary, the paper contributes the following:

\begin{itemize}
    \item We present a compact and low-overhead in-SRAM NTT acceleration design while preserving the generality (i.e., capable of expanding to other crypto kernels) and flexibility (i.e., to easily adjust the bitwidth, polynomial order, and modulus).
    Moreover, these arrays incur minimal hardware modification compared to conventional SRAM arrays (less than 2\%) and can be used for normal cache operations when they are not used as a crypto accelerator.
    
    
    \item To provide a compact and high-throughput computation, we present a bit-parallel data layout, which enables a costless shift for $\sim$50\% of the shift operations in NTT and its inverse (in other words, \#shifts in our bit-parallel design is half of the prior bit-serial solutions). 
    In addition, our design exploits the inspiration from carry-save adder, which eliminates the carry propagation operation in SRAM arrays and enables higher parallelism.

    \item Through simulation, we validate the correctness of the proposed bit-parallel modular multiplication algorithm. 
    Our evaluations reveal that BP-NTT achieves up to an order of magnitude higher throughput-per-area and up to two orders of magnitude higher throughput-per-power compared to the state-of-the-art in-memory, ASIC, and FPGA solutions, on the same technology node and similar parameter settings, thus, making it a low-cost and energy-efficient option for edge devices. 
\end{itemize}

\algrenewcommand\algorithmicrequire{\textbf{Input:}}
\algrenewcommand\algorithmicensure{\textbf{Output:}}

\section{Background}

\subsection{Ring Learning with Errors (R-LWE)}

The R-LWE problem is whether it is possible to create $pk=a*sk+e$ given a public vector $pk$ and $a$ vector, where $a$ is a uniformly sampled vector across $\mathcal{R} \equiv \mathbb{Z}_q[x] /\left\langle x^n+1\right\rangle$, $e$ is a Gaussian distributed error vector of small absolute value, $n$ is a power of 2, $q$ is a prime number, and $sk$ is a secret vector. Consequently, when utilizing a method based on lattice-based cryptography, each encryption/decryption must conduct at least one polynomial multiplication.

\subsection{Number Theoretic Transform (NTT)}

In general, NTT -- which is a generalization of DFT over quotient rings -- utilizes the principles of FFT to lower the complexity of polynomial multiplication from $O(N^2)$ to $O(NlogN)$, where $N$ is the number of polynomial terms. Algorithm \ref{ctntt} illustrates the popular in-place Cooley-Tukey NTT \cite{cooley_algorithm_1965}. The algorithm receives the polynomial $\boldsymbol{a}$ as input and the $n$-th roots on $\mathcal{R}$, and yields the NTT-transformed polynomial $\boldsymbol{a}$ as its output. 
Then, polynomial $\boldsymbol{a}$ can be multiplied with the NTT-transformed polynomial $\boldsymbol{b}$ element-by-element, and the result will be converted from NTT format to standard format using inverse NTT, as $\boldsymbol{a} \boldsymbol{b}=\bera{NTT}^{-1}(\bera{NTT}(\boldsymbol{a})* \bera{NTT}(\boldsymbol{b}))$.


\begin{algorithm}
\small
\algnotext{EndFor}
\caption{Cooley-Tukey-based In-place NTT Algorithm.}
\label{ctntt}
\begin{algorithmic}[1]
\Require $\boldsymbol{a}=(a_{n-1},...,a_0)\in \mathcal{R}$, and pre-computed $n$-th roots $\zeta_{0,...,n-1}$ of unity $w_n$ in $\mathbb{Z}_q$ with bit-reversed order
\Ensure $\boldsymbol{a} = \bera{NTT}(\boldsymbol{a})$ in bit-reversed order

\State $k = 0$
\For{$\texttt{len}=n/2;\ \texttt{len}>0;\ \texttt{len}>>1$}
    \For{$\texttt{idx}=0;\ \texttt{idx}<n;\ \texttt{idx}=j+\texttt{len}$}
        \State $z = \zeta[++k]$ 
            \For{$j=\texttt{idx};\ j<\texttt{idx}+\texttt{len};\ j=j+1$}
                \State $t = z * a[j+\texttt{len}] \mod q$
                \State $a[j+\texttt{len}] = a[j] - t \mod q$
                \State $a[j] = a[j] + t \mod q$
            \EndFor
    \EndFor
\EndFor

\end{algorithmic}
\end{algorithm}
\begin{figure}[htp]
\centerline{\includegraphics[width=3.4in]{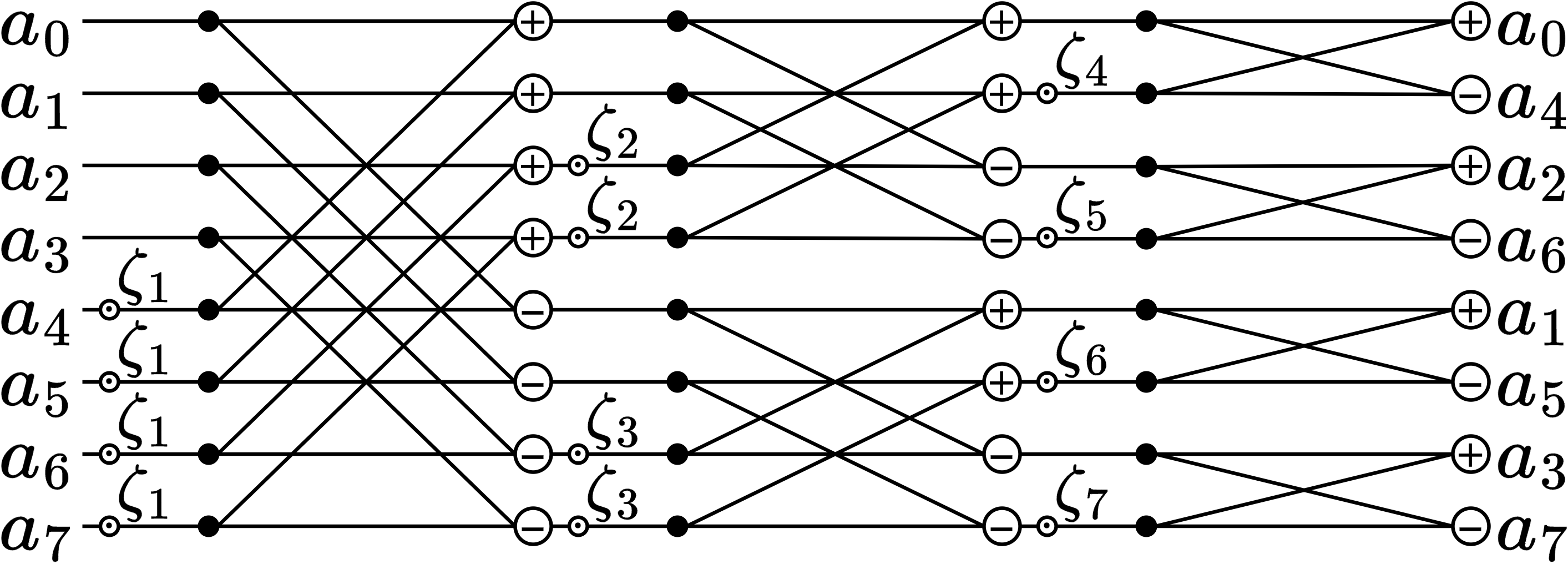}}
\caption{Cooley-Tukey butterfly with 3-stage  communication.}
\vspace{-5mm}
\label{fig:ntt}
\end{figure}

To avoid the costly division required by modular multiplication in NTT, Montgomery multiplication \cite{montgomery_modular_1985} is generally used.
In NTT, the multipliers $\zeta_k$ is pre-computed, which helps avoid the cost of Montgomery domain transformation \cite{mazonka_fast_2022}.
One characteristic of NTT is the intricate data communication between adjacent stages. As depicted in Fig. \ref{fig:ntt}, for an 8-point polynomial, the NTT requires three stages, and the data dependencies between the stages are complicated, resulting in increased data movement, long wires, and higher area overhead when designing ASIC and FPGA-based hardware accelerators. 
For in-memory architecture, this inefficiency is translated to a higher number of shift operations to align the data for bitline computation.  

\subsection{Computing in SRAM}

\begin{wrapfigure}{r}{0.28\textwidth}
\centerline{\includegraphics[width=0.27\textwidth]{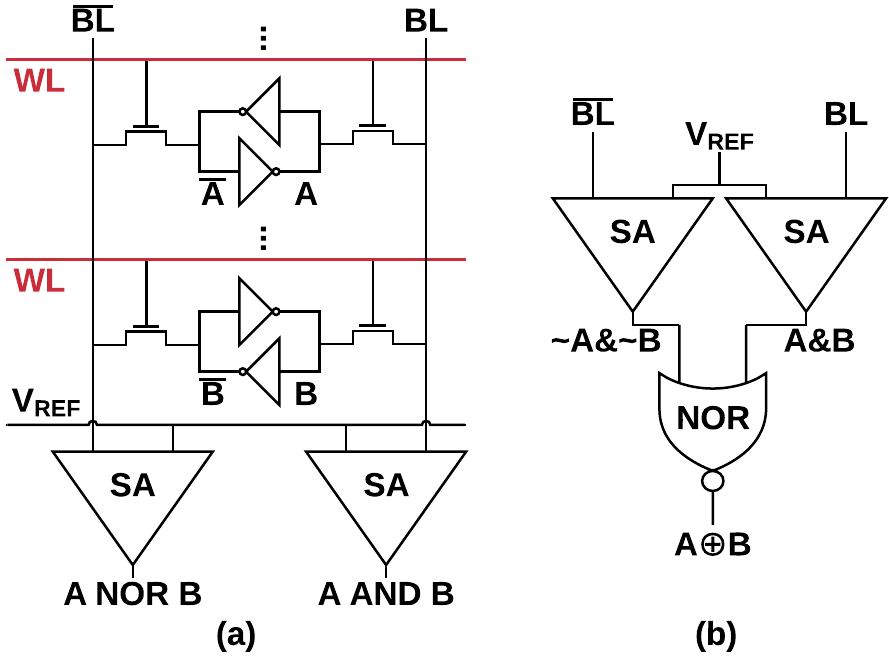}}
\caption{In-SRAM bitline operations.}
\label{fig:bitline}
\end{wrapfigure}
In-SRAM processing is capable of bitline computing \cite{jeloka_28_2016} by activating more than one row in the SRAM subarray. A diagram of the logical operations that can be performed in 6T SRAM is shown in Fig. \ref{fig:bitline}. AND and NOR operations are realized in SRAM with the help of several activated wordlines and sense amplifiers (SAs), as displayed in Fig. \ref{fig:bitline}(a). 
If all the cells in the active rows that are wired to the bitline $BL$ have the value `1', then the SA on the BL will be able to detect a voltage larger than $V_{ref}$, which is an element-wise AND operation.
Only if all the cells in the activated rows connected to the corresponding  $\overline{BL}$  contain `1', the SA on the $\overline{BL}$ senses a voltage greater than $V_{ref}$, which in turn requires that all the cells in the activated rows connected to the corresponding $BL$ contain `0'. This means the SA can do element-wise NOR operations. As can be seen in Fig. \ref{fig:bitline}(b), the XOR operation can be accomplished by combining the capabilities of the logical bit-wise AND and NOR operations.
Cache Automaton \cite{subramaniyan_cache_2017} uses a sense-amplifier cycling mechanism to read out multiple bits in a single time slot, hence reducing input symbol match time. Compute Cache \cite{aga_compute_2017} increases the number of logical operations by modifying the SA architecture based on the NOR, AND, and XOR operations stated in \cite{jeloka_28_2016}. In BP-NTT, we utilize the XOR capability presented in \cite{jeloka_28_2016} and 1-bit shifting (details in Section \ref{sa-design}).

\vspace{-3mm}
\section{Related Work}


\textbf{ASIC/FPGA solutions}: 
Ni et al. \cite{ni_high-performance_2021} reduce lattice-based algorithm latency by accelerating Montgomery multiplication with a three-stage pipeline and systolic array.
Banerjee et al. \cite{banerjee_sapphire_2019} construct a reconfigurable cryptographic processor with a low-power modular arithmetic core to improve hardware overhead and energy efficiency.
Xing et al. \cite{xing_compact_2021} improve latency and hardware complexity by proposing a negative wrapped convolution method to execute polynomial multiplication on FPGAs.
Although the above approaches increase performance, they still suffer from frequent PE-to-memory data movement. Moreover, their designs are limited to a few parameter settings.

\textbf{In-Memory solutions:} 
Nejatollahi et al. \cite{nejatollahi_cryptopim_2020} propose CryptoPIM, a ReRAM-based NTT accelerator that uses a \textit{Shift-Add-based} reduction scheme with the Gentleman-Sande algorithm. However, (1) the fixed interconnection among ReRAM arrays increases hardware overhead and makes it inflexible to support other cryptography kernels, and (2) it only supports a limited number of moduli which limits the flexibility. 
Park et al. \cite{park_rm-ntt_2022} present RM-NTT based on vector-matrix multiplication instead of FFT-like computation to reduce latency. However, it increases memory footprint and energy consumption. In general, ReRAM-based solutions suffer from device variation, which results in reliability problems. In addition, with in-ReRAM computing, the data is located off-chip in plaintext, which makes it vulnerable to memory and bus attacks. 
Li et al. \cite{li_mentt_2022} present MeNTT with a new modular multiplication method and NTT hardware mapping. However, the fixed routing among SRAM arrays and heavy near-memory peripheral circuitry introduces large area/energy overhead and inflexibility.
Unlike prior work, this paper aims to design a flexible, secure, high-performance, low-overhead, and energy-efficient NTT accelerator.

\section{Implementation}
\vspace{-3mm}
\subsection{Overview}

\begin{figure}[tp]
\centerline{\includegraphics[width=2.8in]{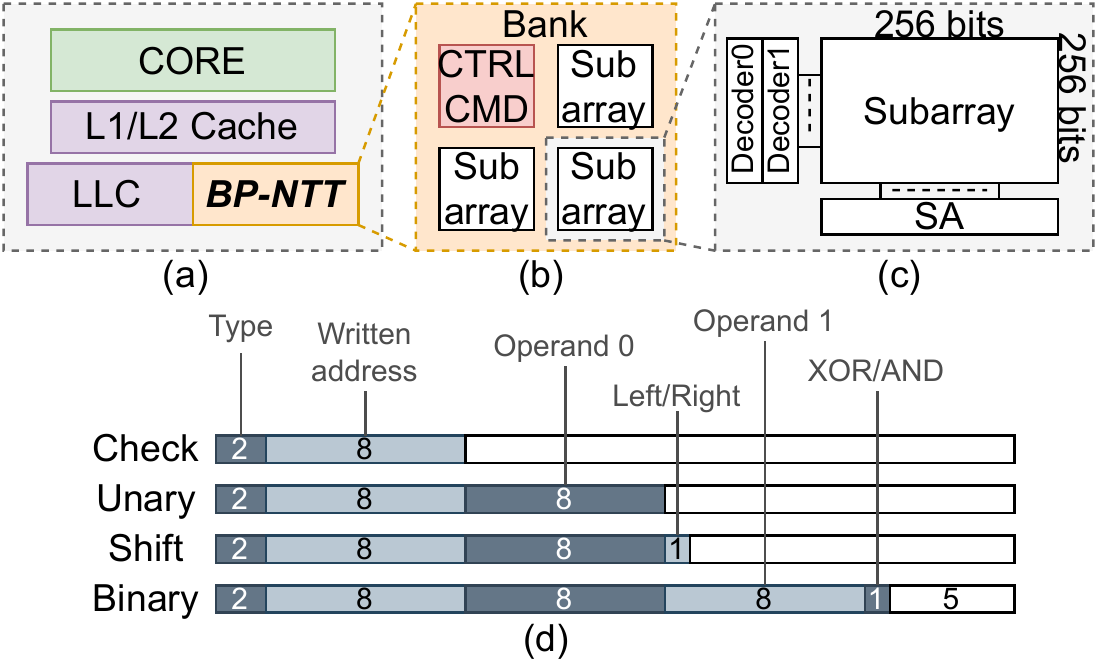}}
\vspace{-2mm}
\caption{(a-c) Hierarchical view of BP-NTT-enabled system. (d) Control signals for different operations in BP-NTT.}
\label{system}
\end{figure}

BP-NTT can be realized either by re-purposing a portion of the L1/L2/L3 cache or by placing a separate in-SRAM accelerator next to the on-chip caches. 
Fig. \ref{system} represents an integration of BP-NTT in the last-level cache (LLC).
Each LLC slice has several SRAM banks, and each bank has usually four subarrays. We repurposed one subarray for memory-mapped command/control instructions (\textit{CTRL/CMD}) and the rest as vector computing units to perform NTT computation, as shown in Fig. \ref{system}(b). Different banks performing the same operations can share \textit{CTRL/CMD} subarray.  
The instruction sets for the BP-NTT are designed and represented in Fig. \ref{system}(d).
Our design incurs minimal hardware changes to the SRAM arrays, i.e., less than 2\% area overhead compared to the conventional SRAM, thus, enabling a low-overhead solution. 

\subsection{Data Organization}

\begin{figure}[tp]
\centerline{\includegraphics[width=2.7in]{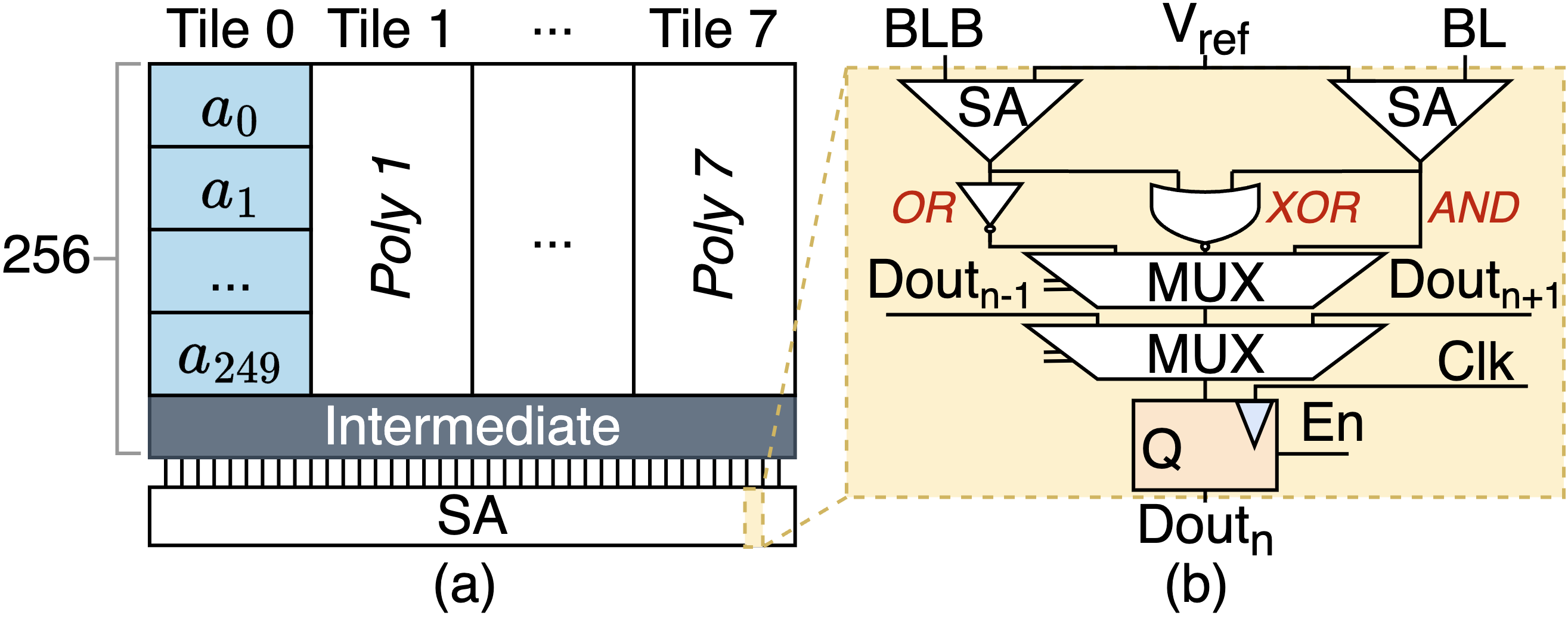}}
\vspace{-2mm}
\caption{(a) Data organization inside one data subarray. (b) The structure of the sense amplifier supporting OR/XOR/AND logic operation and 1-bit left/right shift.}
\vspace{-7mm}
\label{sa}
\end{figure}
In BP-NTT, polynomial coefficients are arranged in distinct rows of the same tile. 
Fig. \ref{sa}(a) shows eight independent polynomials in eight tiles within the same array, with 250 rows for coefficients and 6 rows for intermediate variables. In this configuration, each row in a tile stores 32-bit polynomial coefficients. 
Our design can be easily reconfigured to accommodate $n$ tiles with $\left \lfloor 256/n\right \rfloor$-bit coefficients. When the number of polynomial coefficients is less than the tile's capacity, we can place coefficients from other polynomials in unused rows to save memory. On the other hand, if the number of polynomial coefficients exceeds the tile's capacity, the excess coefficients can be stored in adjacent tiles and merged during computation using the 1-bit shift operation.


Unlike prior methods method \cite{zhang_recryptor_2018}, our method doesn't require shift operations to compute distinct coefficients (i.e., costless shift) thanks to our tile-based data layout that place coefficients in the rows of the same tile so that they can share the bitlines. This means that we can simply select the required coefficient by having their row addresses as the input of the decoders. Although bit-serial alignment doesn't require shifting \cite{eckert_neural_2018} as well, it requires long columns (e.g., 4096 rows for a 128-point 32-bit polynomial), which is very uncommon, especially for resource-constrained edge devices. 


\subsection{Sense Amplifier Design}\label{sa-design}

As shown in Fig. \ref{sa}(b), we modify conventional SRAM SAs to enable in-memory bit-parallel modular multiplication and addition/subtraction in NTT. By using NOR and inverse gates, BP-NTT can perform XOR and OR. A MUX and a latch are introduced to implement 1-bit bidirectional shift operations. 



\subsection{In-memory Bit-parallel Modular Multiplication}

With the aforementioned data arrangement, a polynomial can fit the data compactly into a subarray, which is not possible with a bit-serial design. 
Another benefit of our suggested BP-NTT is its capacity to conduct bit-parallel modular multiplication.
Traditional carry-propagation-based multiplication \cite{eckert_neural_2018,li_mentt_2022} is not suitable for in-memory parallel computation because once carry propagation is introduced, the higher bit must wait for the carry propagation of the lower bits. This hinders maximally exploiting in-memory computing's high parallelism for efficiency.



\begin{algorithm}[tp]
\small
\caption{In-memory Bit-Parallel Modular Multiplication}
\label{alg}
\begin{algorithmic}[1]
\Require $n$-bit $A=(a_{n-1},...,a_0)$, $B=(b_{n-1},...,b_0)$, $M<R=2^n$, where $n>2$ and $M \perp R$
\Ensure $ABR^{-1}\mod{M}$

\State $\texttt{Sum}:=(s_{n-1},...,s_0)=0$ 
\algorithmiccomment{Initialize}
\State $\texttt{Carry}:=(c_{n-1},...,c_0)=0$
\State $P:=\texttt{Sum}+\texttt{Carry}<<1$ \textcolor{purple}{\algorithmiccomment{$P = 0$}}
\For{$i=0$, $n-1$}
    \If{$a_i==1$} \algorithmiccomment{Implicit compare}
        \State $\texttt{\textit{c1}}, \texttt{\textit{s1}} = \{\texttt{Sum}\, \&\, B, \texttt{Sum}\oplus B\}$ 
        \State $\texttt{Carry} << 1$ \hfill <--- \textbf{Oberservation 1}
        \State $\texttt{\textit{c2}}, \texttt{Sum} = \{\texttt{Carry}\, \&\, \texttt{\textit{s1}}, \texttt{Carry}\oplus \texttt{\textit{s1}}\}$
        \State $\texttt{Carry} = \texttt{\textit{c1}}\, |\, \texttt{\textit{c2}}$ \textcolor{purple}{\algorithmiccomment{$P=P+a_iB$}}
    \EndIf
    \State $m = (\bera{LSB(\texttt{Sum})}==1)\ ?\ M\ :\ 0$ \textcolor{purple}{\algorithmiccomment{$m = M \text{ or } 0$}}
    \State $\texttt{\textit{c1}}, \texttt{\textit{s1}} = \{\texttt{Sum}\, \&\, m, \texttt{Sum}\oplus m\}$ 
    \State $\texttt{\textit{s1}} >> 1$ \hfill <--- \textbf{Oberservation 2}
    \State $\texttt{\textit{c2}}, \texttt{\textit{s2}} = \{\texttt{\textit{s1}}\, \&\, \texttt{\textit{c1}}, \texttt{\textit{s1}}\oplus \texttt{\textit{c1}}\}$ 
    \State $\texttt{\textit{c3}}, \texttt{Sum} = \{\texttt{Carry}\, \&\, \texttt{\textit{s2}}, \texttt{Carry}\oplus \texttt{\textit{s2}}\}$
    \State $\texttt{Carry} = \texttt{\textit{c2}}\, |\, \texttt{\textit{c3}}$ \textcolor{purple}{\algorithmiccomment{$P=P+m$; $P>>1$}}
\EndFor

\end{algorithmic}
\end{algorithm}

BP-NTT takes advantage of our proposed in-memory bit-parallel modular multiplication algorithm so that it can perform operations in bit-parallel (e.g., on a 32-bit word) which reduces the latency.
Our proposed algorithm (Algorithm \ref{alg}) is based on Montgomery's algorithm for modular multiplication, which can produce $ABR^{-1}\mod{M}$ for the input polynomials $A$ and $B$, where $R$ is $2^n$, $n$ is the bitwidth, and $M$ is the modulo. 
Notably, although our approach does not directly output $AB\mod{M}$ (same as Montgomery algorithm), the twiddle factors can be pre-computed by multiplying them to $R$ in advance to make the final result ($AB = (AR)BR^{-1} \mod{M}$) as expected without extra conversion.

As shown in Algorithm \ref{alg}, after initialization, our algorithm determines whether to add partial sum $P$ to multiplier $B$ based on the presence of `1' in $A$. In our design, twiddle factor $A$ is hidden in the control commands. Control commands are generated from twiddle factors and stored before starting the NTT computation. For example, if $P = P + B$ is performed on the third iteration, the third bit $a_2$ of twiddle factor $A$ must be $1$. 
In line 11, the least significant bit (LSB) of $Sum$ is used to determine $m$'s value. If $LSB=1$, $m=M$; otherwise, $m=0$. The partial sum is then added to $m$ and shifted to the right. We can get an $n$-bit $Sum$ and $Carry$ after $n$ iterations. 


\begin{table*}
	\centering 
	\small
	\caption{Comparing BP-NTT with state-of-the-art solutions on a 256-point polynomial.}
	\scalebox{0.9}{%
	\begin{tabular}{>{\centering\arraybackslash}p{1.3in}>{\centering\arraybackslash}p{0.6in}>{\centering\arraybackslash}p{0.4in}>{\centering\arraybackslash}p{0.4in}>{\centering\arraybackslash}p{0.4in}>{\centering\arraybackslash}p{0.5in}>{\centering\arraybackslash}p{0.4in}>{\centering\arraybackslash}p{0.3in}>{\centering\arraybackslash}p{0.8in}>{\centering\arraybackslash}p{0.9in}}
		\toprule \midrule
		&  Design &  Coef. Bitwidth &  Max $f$ (MHz) &  Latency ($\mathrm{\mu s}$) &  Tput. (KNTT/s) & Energy (nJ) & Area (mm$ ^2$) & Tput./Area (KNTT/s/mm$ ^2$) & Tput./Power (KNTT/mJ) \\ \midrule
		\textbf{BP-NTT (45nm)}  & In-SRAM  & 16 & 3.8K   & 61.9  & 258.6  & 69.4 & 0.063 & 4.1K & \textbf{230.7}\\ 
		MeNTT (45nm$ ^*$)\cite{li_mentt_2022}   & In-SRAM  & 14 & 218     & 15.9    & 62.8    & 47.8 & 0.173  & 364 & 20.9\\
		CryptoPIM (45nm)\cite{nejatollahi_cryptopim_2020} & ReRAM & 16 & 909  & 68.7 &553.3  &2.6K & 0.152$ \dagger$ & 3.6K & 14.7\\
		RM-NTT (45nm$ ^*$)\cite{park_rm-ntt_2022}  & ReRAM  & 14 & 249 & 0.45 & 2.2K & 602 & 0.289$ \dagger$ & 7.7K & 1.67\\
	 \midrule
        LEIA (45nm$ ^*$)\cite{song_leia_2018}   & ASIC  & 14 & 267     & 0.6    & 1.7K    & 44.1 & 1.77  & 940.6 & 22.7\\
        Sapphire (45nm$ ^*$)\cite{banerjee_sapphire_2019}   & ASIC & 14 & 64     & 20.1    & 49.7    & 236.3 & 0.354  & 140.1 & 4.23\\
		FPGA (45nm$ ^*$)\cite{nejatollahi_exploring_2020}    & FPGA &16 & 164    &24.3   & 41.2  &3061 &- &- &-\\
		CPU\cite{nejatollahi_cryptopim_2020} & x86 & 16 & 2K &85 &11.8 &570K &- &- &-\\ \midrule
		
		\bottomrule
		\addlinespace
	\end{tabular}
	}%
	\label{resulttable}
\raggedright\small \\ \emph{*} Technology nodes are projected to 45nm for an apples-to-apples comparison with BP-NTT.\\
\emph{$\dagger$} These solutions do not provide the area consumption. For the sake of comparison, we optimistically estimate their area overhead based on the area of memory subarrays used in these designs (i.e., we ignore the peripheral overhead).
\vspace{-4mm}
\end{table*}

The proposed algorithm uses two variables, $Sum$ and $Carry$, to represent the sum of two addends to enable in-memory bit-parallel modular multiplication. Inspired by carry-save adder design \cite{noauthor_carry-save_2021}, $Sum$ and $Carry$ are derived directly using bitwise operations (AND, XOR, OR) and kept distinct throughout the computation.


\begin{figure}[tp]
\centerline{\includegraphics[width=3in]{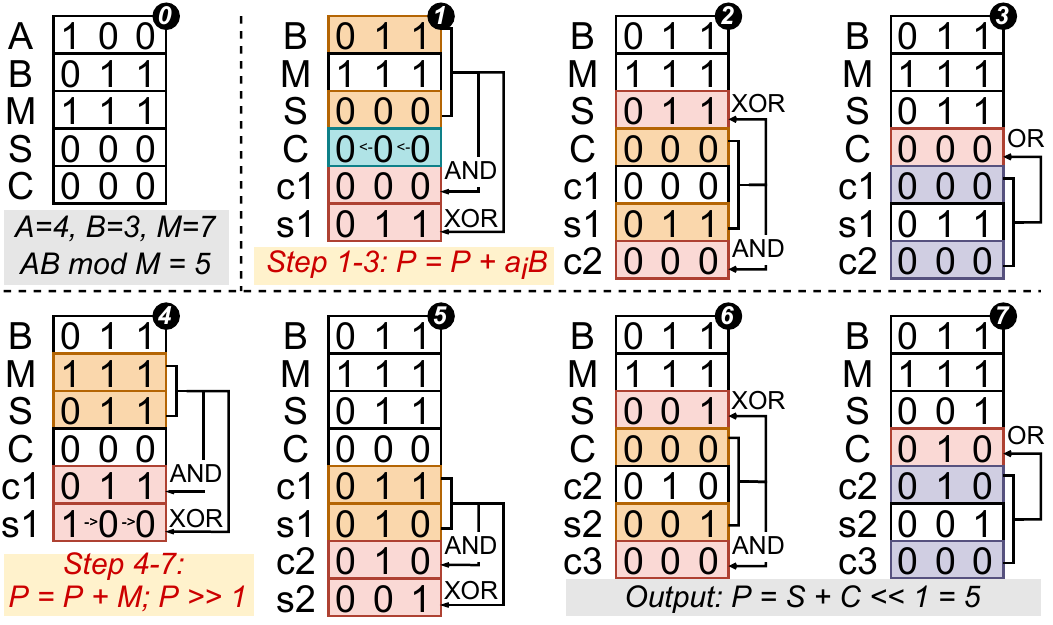}}
\caption{A 3-bit example of the proposed in-memory bit-parallel modular multiplication.}
\vspace{-5mm}
\label{eg}
\end{figure}

The number of bits in the vanilla algorithm can be reduced from $n + 1$ to $n$ based on two observations: (1) the highest bit of the $Carry$ is always $0$ after each iteration; and (2) the lowest bit of the result in $P=P+M$ is always $0$. Thus by shifting $Carry$ 1-bit to the left (line 7) and $s1$ 1-bit to the right (line 13), all the computations can be done within $n$ columns, which is crucial for exploiting parallelism in in-SRAM computing. A 256-column SRAM array with 32-bit operands can only perform seven modular multiplications in parallel if $n + 1$ bits (33 columns) is required, whose throughput is 12.5\% worse than our proposed method.




Fig. \ref{eg} depicts an example of bit-parallel modular multiplication in SRAM. Inputs are $A=4$, $B=3$, and $M=7$. We use $A$ to directly represent $AR$ because $A = AR \mod M$. Due to the lowest two bits of $A$, $P$ remains $0$ after two iterations. In step \Circled{\scriptsize{\textit{\textbf{1}}}} of the third iteration, $B$ and $Sum$ are bit-parallelly ANDed and XORed to generate \texttt{c1} and \texttt{s1}. $Carry$ is shifted to the left by one bit to align with $Sum$ for subsequent addition operations. Step \Circled{\scriptsize{\textit{\textbf{2}}}} performs AND and XOR on $Carry$ and \texttt{s1} to produce \texttt{c2} and $Sum$. In step \Circled{\scriptsize{\textit{\textbf{3}}}}, \texttt{c1} and \texttt{c2} are ORed to get $Carry$. In step \Circled{\scriptsize{\textit{\textbf{4}}}}, $M$ and $Sum$ produce \texttt{c1} and \texttt{s1}. For addition, \texttt{s1} is shifted to the right by one bit. Steps \Circled{\scriptsize{\textit{\textbf{5}}}}-\Circled{\scriptsize{\textit{\textbf{7}}}} are the same as \Circled{\scriptsize{\textit{\textbf{1}}}}-\Circled{\scriptsize{\textit{\textbf{3}}}}. Finally, we can obtain the result $P = 001 + 010 < < 1 = 5$.


\vspace{-1mm}
\subsection{Implicit Shift in NTT}
Inefficient hardware and complex data dependencies make NTT acceleration challenging. The product of modular multiplication should be added to or subtracted from polynomial coefficients (lines 7-8 in Algorithm \ref{ctntt}), which is difficult with in-memory computing due to shifting overhead. The implicit shift design prevents word shifting. Thanks to our data organization, which places all polynomial coefficients in the same tile, we can perform bitline computing by implicitly selecting the corresponding row of operands, thus, avoiding shift operations. If we want to add the product of modular multiplication with the $k$-th polynomial coefficient, we only need to activate the rows where the product and the $n$-th coefficient are located, to enable bitline computing and perform the standard addition with 1-bit shifts. 
As a result, compared to MeNTT \cite{li_mentt_2022}, our BP-NTT consumes less energy due to the reduced shifting overhead.

\section{Evaluation}\label{sec:evaluation}

\subsection{Evaluation Methodology}

In this section, we analyze the latency, throughput, area, energy, throughput-per-area, and throughput-per-power of BP-NTT performing NTT for 256-point polynomials with 14-bit and 16-bit coefficients (as they are the most common parameter choices in PQC algorithms \cite{bos_crystals-kyber_2018,ducas_crystals-dilithium_2018}), and compare them to existing ASICs, FPGAs, and in-memory designs, as shown in Table \ref{resulttable}. 
The correctness of the proposed bit-parallel modular multiplication has been validated for various bitwidths.
The array size of BP-NTT is 256$\times$256 following the ARM Cortex-M0+ microcontroller \cite{noauthor_lpc1225fbd48arm_nodate}. 
We then investigate BP-NTT performance on different parameter configurations.
PyMTL3\cite{jiang_pymtl3_2020} and OpenRAM\cite{guthaus_openram_2016} are utilized to construct SRAM arrays, and Synopsys Design Compiler and Cadence Innovus are utilized to calculate read and write latency and area consumption, respectively. 
Additionally, due to the lack of area consumption analysis in CryptoPIM \cite{nejatollahi_cryptopim_2020} and RM-NTT \cite{park_rm-ntt_2022}, we utilize the Destiny simulator \cite{poremba_destiny_2015} to optimistically estimate only the subarray areas, and we do not account for their complex peripheral circuitry.

\subsection{Memory Footprint and Area Comparison}

\begin{figure}[tp]
\centerline{\includegraphics[width=3in]{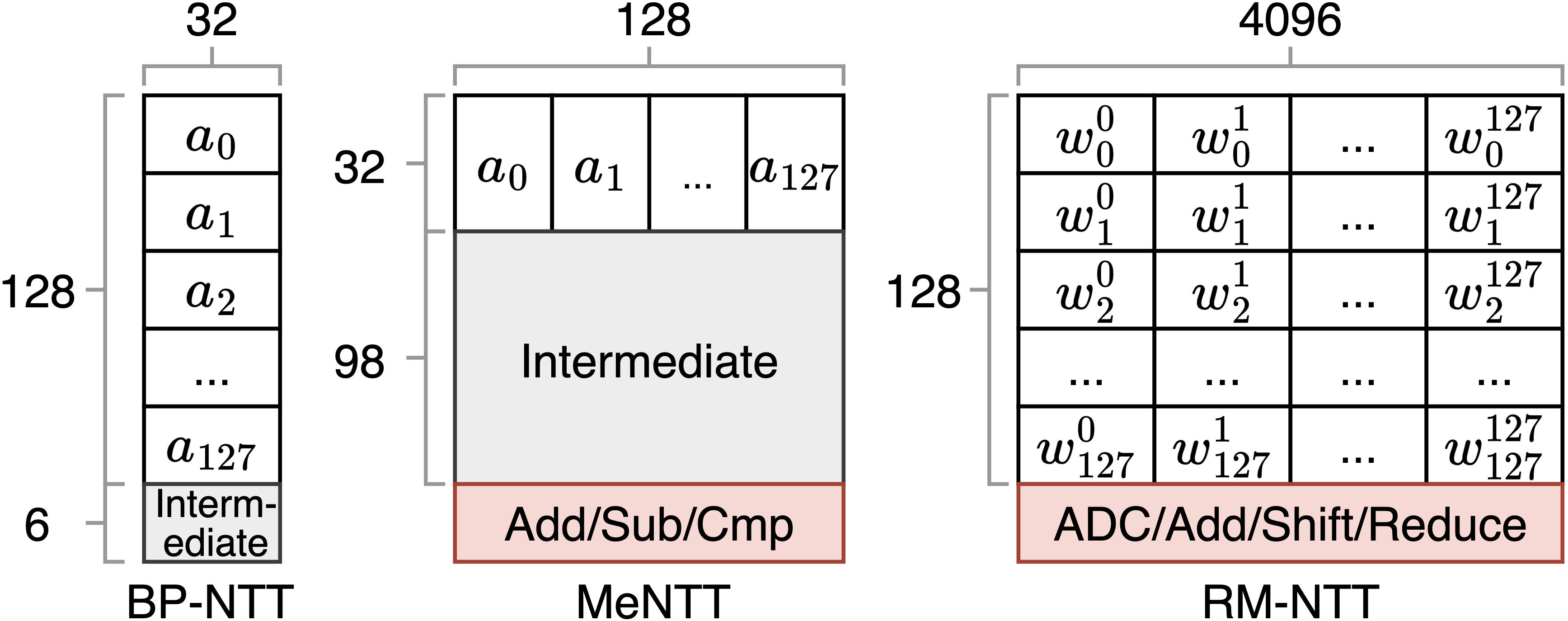}}
\caption{Comparison of different in-memory designs for NTT on a 32-bit 128-point polynomial.}
\vspace{-4mm}
\label{cmp}
\end{figure}

BP-NTT occupies the least amount of memory footprint and introduces the least amount of hardware overhead compared to other in-memory designs. Fig. \ref{cmp} shows the data layout of BP-NTT, MeNTT, and RM-NTT for computing NTT with 32-bit, 128-point polynomial.
BP-NTT requires only 4288 SRAM cells (i.e., 134 rows and 32 columns), however, MeNTT and RM-NTT require 16,640 (130 rows and 128 columns) and 524,288 (128 rows and 4096 columns) ReRAM cells, respectively. In addition, BP-NTT is capable of performing all NTT-related operations within an SRAM subarray, thanks to our compact data organization and the proposed bit-parallel modular multiplication. However, MeNTT requires peripheral circuitry for addition, subtraction, and comparison operations, and RM-NTT requires additional shift and reduction modules, resulting in a large area and energy consumption overhead.


As shown in Table \ref{resulttable}, BP-NTT provides at least 2.4$\times$-4.6$\times$ lower area overhead compared to the state-of-the-art in-memory designs. This is because their data arrangement and operation mode (i.e., bit-serial computing in MeNTT and vector-matrix-based NTT in RM-NTT) necessitates more memory capacity to support NTT acceleration.  





\subsection{Throughput-per-Area \& Throughput-per-Power}


BP-NTT is a design that exhibits the highest throughput-per-power (TP) when compared to other designs. This is due to its energy efficiency, high performance, and minimal overhead, rendering it a suitable candidate for IoT devices. The proposed bit-parallel modular multiplication and data organization enable costless shift operations, contributing to BP-NTT's superior performance. In addition, BP-NTT provides excellent throughput-per-area (TA) by making minimal modifications to existing standard arrays, and it only requires one array to perform several NTTs with different parameters, resulting in a low-overhead design. In general, ReRAM-based in-memory designs offer higher throughput due to their efficient analog computing capabilities. However, their computing schemes, including cascaded pipeline and vector-matrix-based NTT, require a large number of arrays and memory cells, resulting in high energy consumption. BP-NTT outperforms ASIC/FPGA designs by up to 30$\times$ and 50$\times$ in TA and TP metrics, respectively, due to the latter's frequent data movement.

\subsection{Latency Comparison}
The higher latency of BP-NTT compared to ASIC/FPGA designs is due to the fact that ASIC/FPGA can fine-tune algorithms so that unnecessary operations on the critical path can be minimized for certain algorithms and parameter settings. 
On the other hand, hardware generalization and flexibility are sacrificed.
In-memory designs have lower latency than BP-NTT because to handle inefficient modular multiplication and complex inter-stage data communication, either they target a specific modulus and fixed inter-array connections, or they introduce a large number of combinational logic circuits (comparators, adders, subtractors, and even Montgomery reduction module), thereby, sacrificing flexibility and adding unnecessary area overhead. 
In contrast, BP-NTT can accomplish all NTT operations within one subarray with minimal modifications to the SA and without any extra dedicated modules.

\subsection{Flexibility Analysis}
\begin{figure}[htp]
\centering
\subfloat[Polynomial order = 256]{%
  \includegraphics[width=2.4in]{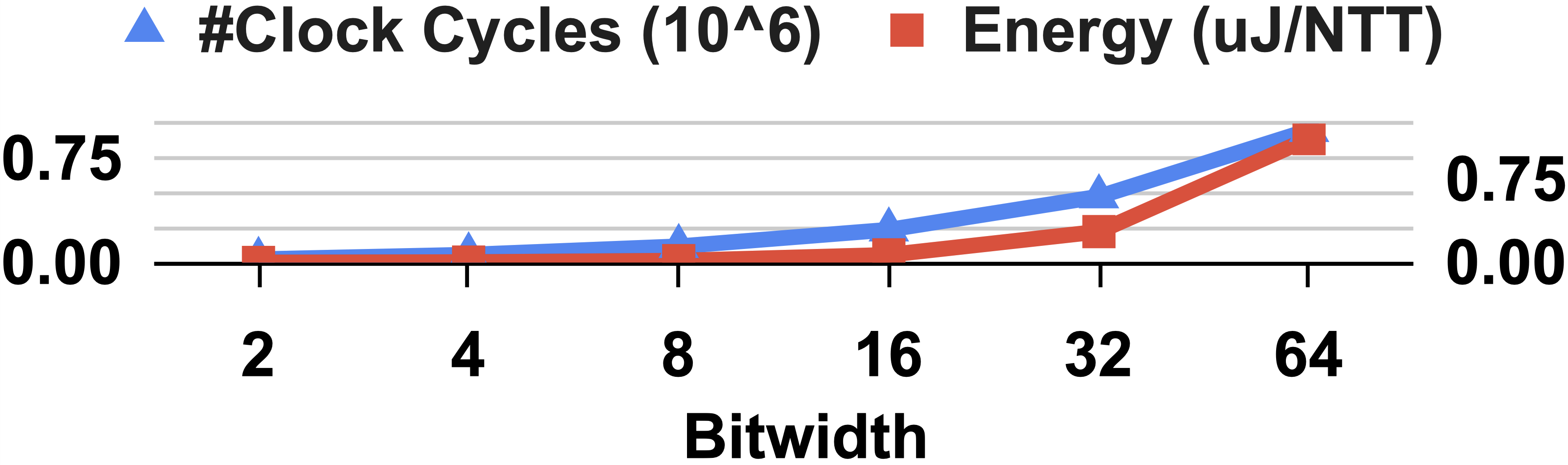}%
}

\subfloat[Bitwidth = 16]{%
  \includegraphics[width=2.2in]{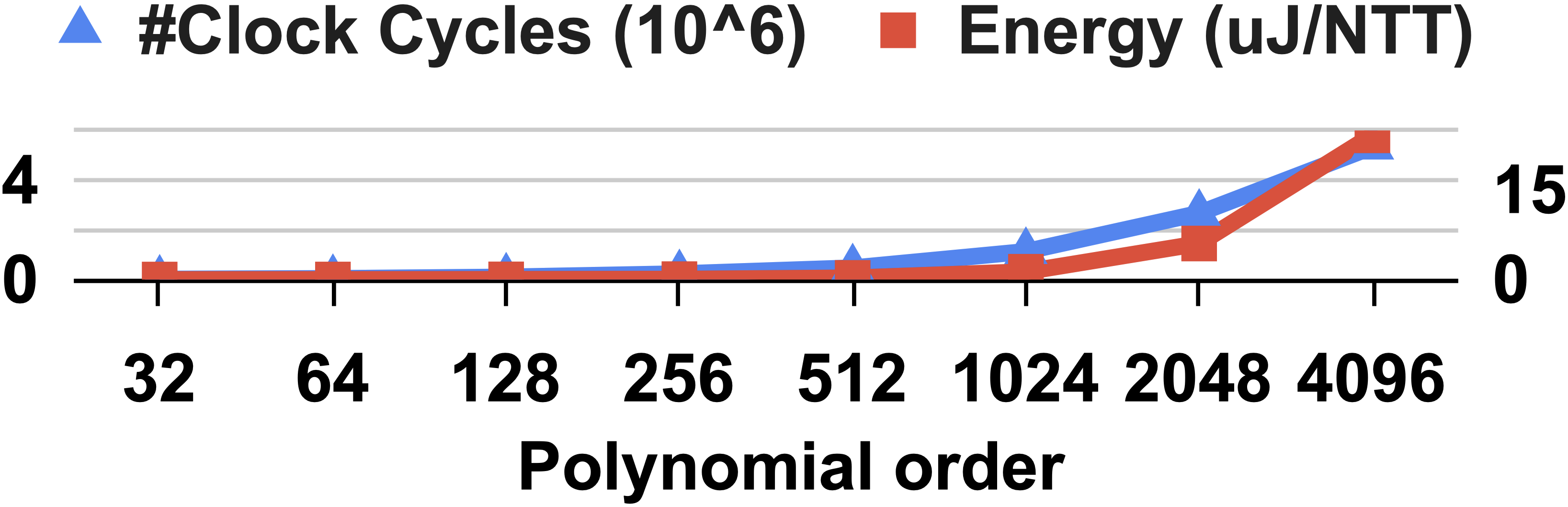}%
}
\caption{Performance and energy consumption of BP-NTT (a) with different bitwidths when polynomial order is 256, (b) with different polynomial orders when bitwidth is 16.}
\vspace{-5mm}
\label{line}
\end{figure}

The BP-NTT is flexible to support NTT calculations with different parameters.
Figure \ref{line}(a) shows the clock count and energy overhead for a 256$\times$256 BP-NTT design plus 6 rows for intermediate data using 2-bit to 64-bit coefficients with a polynomial order of 256.
As the bitwidth increases, both the clock count and energy overhead grow.
The reason for the steeper increase in energy overhead is that as bitwidth grows, the number of NTTs that can be computed in parallel in a fixed-size subarray decreases.


Fig. \ref{line}(b) shows the number of clock cycles and energy consumption for performing NTT on BP-NTT with different polynomial orders at a bitwidth of 16.
The steeper increase in clock count and energy consumption compared to Fig. \ref{line}(a) is because as the polynomial order rises, not only does the number of NTTs that can be computed in parallel in the subarray decrease but also the additional shift overhead in a single NTT is introduced.
However, these additional overheads can be effectively avoided by using the larger subarray or interconnection of multiple subarrays.

\section{Conclusion}
In this paper, we present BP-NTT, an in-SRAM architecture for flexible, secure, high-performance, low-overhead, energy-efficient acceleration of NTT. By utilizing bit-parallel modular multiplication, we take advantage of the parallelism of in-SRAM computing to enhance performance. By incorporating our proposed data organization, the majority of shift operations can be eliminated with minimal hardware modifications. Our evaluation results indicate that BP-NTT can achieve a significant improvement in throughput-per-power (up to 138$\times$) over the latest ASIC/FPGA and in-memory designs.

\vspace{-1mm}
\section{ACKNOWLEDGMENTS}
This work is funded, in part, by the Hellman Fellowship from the University of California, NSF \#2127780, Semiconductor Research Corporation, and Office of Naval Research.
\vspace{-5mm}

\bibliographystyle{IEEEtran.bst}
\normalsize{
\bibliography{IEEEabrv.bib,references.bib}
}

\end{document}